\documentclass[showpacs, pra, twocolumn, preprintnumbers, amsmath, amssymb, superscriptaddress, aps]{revtex4-2}
\usepackage{color}
\usepackage{amsmath,amssymb}
\usepackage{pifont}
\usepackage{amssymb}  
\usepackage{bbold}
\usepackage{float}
\usepackage{subfloat}
\usepackage[caption=false]{subfig}
\usepackage{tikz}
\usepackage{makecell}
\usepackage{subfig}
\usepackage{pifont}   
\usepackage{graphicx} 
\graphicspath{{Figures/}}
\usepackage{dcolumn}  
\usepackage{bm}       
\usepackage{multirow} 
\usepackage[colorlinks]{hyperref}
\usepackage{placeins}
\usepackage{mathtools}
\usepackage{appendix}

\begin{document}

        \title{
        Effect of a perpendicular magnetic field on	bilayer graphene under dual gating
}
        \date{\today}
        \author{Mouhamadou Hassane Saley }
        \email{hassanesaley.m@ucd.ac.ma}
        \affiliation{Laboratory of Theoretical Physics, Faculty of Sciences, Choua\"ib Doukkali University, PO Box 20, 24000 El Jadida, Morocco}
        \author{ Abderrahim El Mouhafid}
        \email{elmouhafid.a@ucd.ac.ma}
        \affiliation{Laboratory of Theoretical Physics, Faculty of Sciences, Choua\"ib Doukkali University, PO Box 20, 24000 El Jadida, Morocco}
        \author{Ahmed Jellal}
        \email{a.jellal@ucd.ac.ma}
        \affiliation{Laboratory of Theoretical Physics, Faculty of Sciences, Choua\"ib Doukkali University, PO Box 20, 24000 El Jadida, Morocco}
        \affiliation{Canadian Quantum Research Center, 204-3002 32 Ave Vernon,  BC V1T 2L7, Canada}

        
        \begin{abstract} 
        	
        	By studying the impact of a perpendicular magnetic field $B$ on AB-bilayer graphene (AB-BLG) under dual gating, we yield several key findings for the ballistic transport of gate $U_\infty$. Firstly, we discover that the presence of $B$ leads to a decrease in transmission. At a high value of $B$, we notice the occurrence of anti-Klein tunneling over a significant area. Secondly, in contrast to the results reported in the literature, where high peaks were found with an increasing in-plane pseudomagnetic field applied to AB-BLG, we find a decrease in conductivity as $B$ increases. However, it is worth noting that in both cases, the number of oscillations decreases compared to the result in the study where no magnetic field was present $(B = 0)$. Thirdly, at the neutrality point, we demonstrate that the conductivity decreases and eventually reaches zero for a high value of $B$, which contrasts with the result that the conductivity remains unchanged regardless of the value taken by the in-plane field. Finally, we consider the diffusive transport with gate $U_\infty = 0.2 \gamma_1$ and observe two scenarios. The amplitude of conductivity oscillations increases with $B$ for energy  $E$ less than $U_\infty$ but decreases in the opposite case $E>U_\infty$. 


\end{abstract}          

\maketitle

\section{Introduction}

The stacking of multiple layers of graphene  \cite{graphene} can lead to the formation of new materials with unique properties \cite{McCann,Peeters,Guinea,Nilsson,Jung,Mekkaoui,Mouhafid,Mouhamad}. We cite AB bilayer graphene (AB-BLG), which stands out due to its ability to develop and regulate a gap through the application of an external electric field \cite{McCann,Zasada,Li,Van,Hassane,Saley,Nadia}. In contrast to MLG, AB-BLG exhibits a parabolic dispersion with two bands that converge at the Dirac points \cite{Van,Hassane,Saley,Nadia,Abergel,Snyman} in addition to two bands separated from the Dirac points by an energy equivalent to the interlayer coupling $\gamma_{1}\simeq 0.4$eV \cite{Li}. On the other hand, AA-BLG demonstrates a linear energy dispersion akin to MLG, featuring two Dirac cones shifted by an energy $\gamma\simeq0.2$eV \cite{Lobato,Sanderson,Abdelhadi}. Furthermore, AB-BLG displays anti-Klein tunneling \cite{Peeters,Van,Hassane,Nadia,Saley}, while AA-BLG is characterized by perfect electron transmission, akin to MLG \cite{Sanderson}. 
Also, AB-BLG shows a distinct quantum Hall effect, featuring a Berry phase of $2\pi$ \cite{Schedin}, and setting it apart from MLG.

Recent research in the field of ballistic transport has presented evidence indicating that the conductivity at the neutrality point in AB-BLG remains unchanged when an in-plane pseudomagnetic field is introduced \cite{Abdullah}. This observation aligns with the conclusions drawn in a previous study that focused on the behavior of undoped AB-BLG \cite{Snyman}. These findings prompted us to explore the influence of a perpendicular magnetic field $B$ on ballistic transmission in AB-BLG. Our study demonstrates that the presence of $B$ results in a reduction in transmission and the number of resonances. Notably, we observe the occurrence of anti-Klein tunneling over a large area at a specific value $B=14.44$T. Additionally, as $B$ increases, both the number and amplitude of oscillations in conductivity decrease. Furthermore, at the neutrality point, the conductivity reaches zero at $B=14.44$T. Regarding the diffusive process, we choose $U_\infty=0.2 \gamma_1$ and show that the amplitude of conductivity oscillations increases when $B$ increases in the case $E<U_{\infty}$, while it decreases in the opposite scenario.

The paper is organized as follows. In Sec. \ref{TTMM}, we introduce a theoretical model describing the present system and determine the associated solutions of the energy spectrum. Sec. \ref{Trans} focuses on determining the transmission probability using continuity conditions and the transfer matrix method. In Sec. \ref{RRDD}, we numerically present and discuss the obtained results. Finally, we provide a summary of our findings in Sec. \ref{CC}.

\section{Theory and methods}\label{TTMM}
\begin{figure}[h]
        \centering
        \includegraphics[width=3 in]{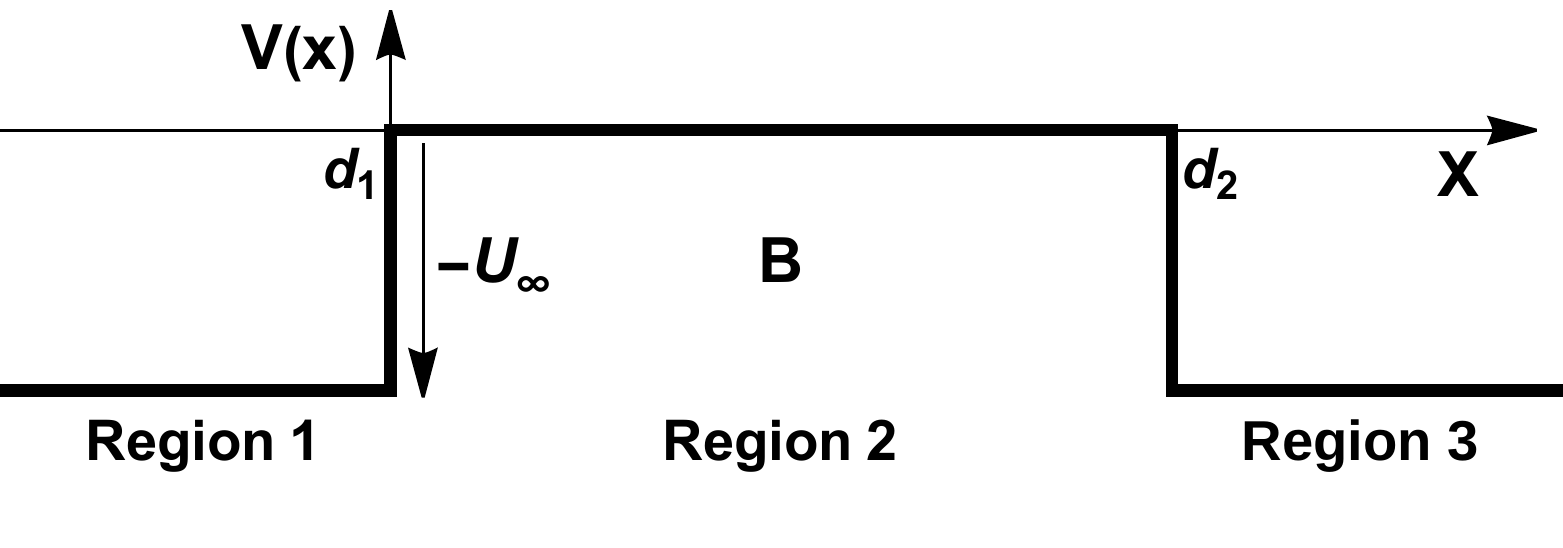} \caption{The profile of a potential applied to graphene subjected to a perpendicular magnetic field in the central region (region 2).}\label{shematicstep}
\end{figure}
We consider AB-BLG submitted to a perpendicular magnetic field and dual gates, as depicted in Fig. \ref{shematicstep}. Such gates manifest themselves through the application of a potential to generate a strip of weakly doped (length $L=d_2-d_1$, width $W$) and higher-doped contact regions for $x<d_1$ and $x>d_2$. Otherwise, we write the potential as
 \begin{equation} 
        U_{j}= \begin{cases}-U_{\infty}, & \text{if} ~ x<d_1,   x>d_2\\
                0, & \text{if}~ d_1<x<d_2,\end{cases}
        \label{Eq1}
\end{equation}
and $j=1,2,3$ labels the three regions. The  magnetic field applied in region 2 takes the form
\begin{equation}
        B(x, y)=B \Theta\left[\left(d_1-x\right)\left(d_2-x\right)\right], 
\end{equation}
with $B$ is a constant and $\Theta(x)$ is the step function. The corresponding vector potential  in the Landau gauge is given by
\begin{equation}
 A_{y}(x)=\frac{\hbar}{e l_{B}^{2}} \begin{cases}
                        d_1, & \text{if} \quad x<d_1, \\
                        x, & \text{if}  \quad d_1<x<d_2, \\
                        d_2, & \text{if} \quad x>d_2, 
              \end{cases}      \label{Eq2}
\end{equation} 
and  $l_B=\sqrt{\hbar/eB}$ is the magnetic length. 
To perform a theoretical analysis of the present system, we formulate a Hamiltonian that incorporates all the mentioned constraints. It can be expressed as follows \cite{Snyman,Van,Redouani2015}
\begin{equation}
        H_{j}= \begin{pmatrix}
                U_{j} & v_{F} \pi^{\dagger} &0  &0 \\
                v_{F} \pi &     U_{j}& \gamma_{1} & 0  \\
                0 & \gamma_{1} &        U_{j} & v_{F} \pi^{\dagger} \\
                0 & 0  &        v_{F} \pi &     U_{j}
        \end{pmatrix},
        \label{Eq3}
\end{equation}  
where $v_{F}=10^{6}$ m/s is the Fermi velocity and $\pi=p_{x}+i( p_{y}+eA_{y}(x))$ is the in-plan momentum. The Hamiltonian \eqref{Eq3} operates on the spinor of four components, i.e., $\Psi_{j}(x, y)=\left(\psi_{A_{1}j}, \psi_{B_{1}j}, \psi_{A_{2}j}, \psi_{B_{2}j}\right)^{\dagger}$, with the symbol $\dagger$ standing for the transpose row vector. Since the momentum is conserved in the $y$-direction, the spinor can be decomposed into
\begin{equation}
        \Psi_{j}(x, y)=e^{i k_{y}y}\left(\psi_{A_{1}j}, \psi_{B_{1}j}, \psi_{A_{2}j}, \psi_{B_{2}j}\right)^{\dagger}.
        \label{Eq4}
\end{equation}
Now, we combine (\ref{Eq3}) with (\ref{Eq4}) into the eigenvalue equation $H_{j} \Psi_{j}=E_{j} \Psi_{j}$ to arrive at
\begin{subequations}
\begin{align}
                -i \sqrt{2} \vartheta_{0} a \psi_{B_{1}j}&=\varepsilon_{j}\psi_{A_{1}j},\label{Eq5a}
                \\
                i \sqrt{2} \vartheta_{0} a^{\dagger} \psi_{A_{1}j}&=\varepsilon_{j}\psi_{B_{1}j}-\gamma_{1}\psi_{A_{2}j}, \label{Eq5b}
                \\
                -i \sqrt{2} \vartheta_{0} a \psi_{B_{2}j}&=\varepsilon_{j}\psi_{A_{2}j}-\gamma_{1}\psi_{B_{1}j}, \label{Eq5c} 
                \\
                i \sqrt{2} \vartheta_{0} a^{ \dagger} \psi_{A_{2}j}&=\varepsilon_{j}\psi_{B_{2}j},
                \label{Eq5d}
        \end{align}
\end{subequations}
where the annihilation \(a\) and  creation  \(a^{\dagger}\) operators are represented by 
\begin{subequations}
        \begin{align}
               & a=\frac{l_{B}}{\sqrt{2}}\left(\partial_{x}+k_{y} +\frac{eA_{y}(x)}{\hbar}\right),
                        \label{Eq6a}\\
                &a^{\dagger}=\frac{l_{B}}{\sqrt{2}}\left(-\partial_{x}+k_{y} +\frac{eA_{y}(x)}{\hbar}\right),
                \label{Eq6b}
        \end{align}
\end{subequations} 
which  satisfy the commutation relation \([a, a^{\dagger}] = \mathbb{I}\). Here, we have set the energy   $\varepsilon_{j}=E_{j}-U_{j}$, with   $\vartheta_{0}=\frac{ \hbar v_{F}}{l_B}$ denoting an energy scale. From Eqs. (\ref{Eq5a}-\ref{Eq5d}), we can, for instance, establish a four order differential equation for  the wave function $\psi_{B_{1j}}$
\begin{align}
        \left[ 2\vartheta_{0}^{2}aa^{\dagger}-\varepsilon_{j}^{2}\right]\left[ 2\vartheta_{0}^{2}a^{\dagger}a-\varepsilon_{j}^{2}\right]\psi_{B_{1j}}  =\gamma_{1}^{2}\varepsilon_{j}^{2} \psi_{B_{1j}}.
                        \label{Eq7}
\end{align}
                
In region 2 ($d_1<x<d_2$), where $A_{y}(x)=\frac{\hbar x}{e l_{B}^{2}}$ and $U_\infty=0$, we solve Eq. (\ref{Eq7}) to derive $\psi_{B_{1}}$ and the result is employed in Eqs. (\ref{Eq5a}-\ref{Eq5d}) to determine the other components of the spinor. Thereafter, the solution of the spinor can be expressed in matrix form
\begin{equation}
        \Psi_2(x, y)=\mathcal{L}_2 \mathcal{M}_2(x) \mathcal{C}_2  e^{i k_y y},
        \label{Eq8}
\end{equation}
where $\mathcal{C}_2 = (\alpha_1,\beta_1,\alpha_2,\beta_2)^\dagger$ is a column matrix of four coefficients, $\mathcal{L}_2=\mathbb{I}_4$ is a $4 \times 4$ identity matrix, and $ \mathcal{M_{\text{2}}}(x)$ is given by
\begin{widetext}
\begin{equation}
        \mathcal{M_{\text{2}}}(x)=
        \begin{pmatrix}
                        \eta \lambda_{+} \chi_{+,-1}^{+} & \eta^* \lambda_{+} \chi_{-,-1}^{+} & \eta \lambda_{-} \chi_{+,-1}^{-} & \eta^* \lambda_{-} \chi_{-,-1}^{-} \\
                        \chi_{+,0}^{+} & \chi_{-,0}^{+} & \chi_{+,0}^{-} & \chi_{-,0}^{-} \\
                        \zeta_{+} \chi_{+,0}^{+} & \zeta_{+} \chi_{-,0}^{+} & \zeta_{-} \chi_{+,0}^{-} & \zeta_{-} \chi_{-,0}^{-}  \\
                        \eta^* \zeta_{+} \chi_{+,1}^{+} & \eta \zeta_{+} \chi_{-,1}^{+}  & \eta^* \zeta_{-} \chi_{+,1}^{-} & \eta \zeta_{-} \chi_{-,1}^{-} 
               \end{pmatrix}.
\label{Eq13}
\end{equation}
\end{widetext}
Here, we have introduced the Weber parabolic cylindrical function $\chi^{\tau}_{\pm, s}= D\left[\lambda_{\tau}\pm s,\pm z \right]$  of argument $z=\sqrt{2}(\frac{x}{l_{B}}+k_{y}l_{B})$, $ s=-1, 0, 1 $,   $\tau=\pm$, and we have set the following parameters
\begin{subequations}
        \begin{align}
&\lambda_{\tau}=-\frac{1}{2}+\frac{E^{2}}{2\vartheta_{0}^{2}}+\tau\frac{\sqrt{\vartheta_{0}^{4}+\gamma_{1}^{2}E^2}}{2\vartheta_{0}^{2}}, \label{E10a}\\ 
&
\eta=\frac{-i\sqrt{2}\vartheta_{0}}{E}, \quad      \zeta_{\tau}=\frac{E^2-2\vartheta_{0}^{2}\lambda_{\tau}}{\gamma_{1}E}.\label{E10b}            \end{align}
\end{subequations}
As a result, from Eq. (\ref{E10a}), we derive the energy spectrum 
\begin{equation}
        E^\tau_{n\pm}=\tau \sqrt{(2 n+1) \vartheta_0^2+\frac{\gamma_1^2}{2} \pm \sqrt{(2 n+1) \vartheta_0^2 \gamma_1^2+\vartheta_0^4+\frac{\gamma_1^4}{4}}},
\end{equation}
and  $n$ is an integer value, indexing the Landau levels.

Recall that in regions 1 ($x<d_1$) and  3 ($x>d_2$), the vector potential $A_{y}(x)=\frac{ \hbar }{e l_{B}^{2}}d_{j}$ is constant where
\begin{equation}
d_{j}=\left\{\begin{array}{lll}
                d_1, & \text { if } & x<d_1, \\
                d_2, & \text { if } & x>d_2.
        \end{array}\right.
        \label{Eq23}
        \end{equation}
Upon solving Eq. (\ref{Eq7}) for $\psi_{B_{1}}$ and substituting the obtained result into Eqs. (\ref{Eq5a}-\ref{Eq5d}), a general solution of the spinor is derived in matrix form
\begin{equation}
        \Psi_j(x, y)=\mathcal{L}_j \mathcal{M}_j(x) \mathcal{C}_j e^{i k_y y},
        \label{Eq24}
\end{equation}
where $\mathcal{L_{\text{$j$}}}$, $\mathcal{M_{\text{$j$}}}(x)$, and  $\mathcal{C}_j$ are given, respectively, by 
\begin{align}
    &    \mathcal{L}_j=\begin{pmatrix}
                        f_{-j}^{+} & -f_{+j}^{+} & f_{-j}^{-} & -f_{+j}^{-} \\
                        1 & 1 & 1 & 1 \\
                        -1 & -1 & 1 &1 \\
                        - f_{+j}^{+} &  f_{-j}^{+} &  f_{+j}^{-} & - f_{-j}^{-}
                \end{pmatrix},
\label{Eq25}
\\
&
       \mathcal{M}_j(x)=\text{diag}\left[
                        e^{i k_{j}^+ x}, e^{-i k_{j}^+ x}, e^{i k_{j}^- x}, e^{-i k_{j}^- x}
               \right],
\label{Eq26}\\
& \mathcal{C}_j=(a_j,b_j,c_j,e_j)^\dagger\label{Eq266}
\end{align}
and we have defined the parameters
\begin{equation}
        f_{\pm}^\tau=\hbar v_F \frac{k_j^\tau \pm i\left(k_y+\frac{d_j}{l_B^2}\right)}{\varepsilon_j}.
        \label{Eq27}
 \end{equation}
 We show that the associated energies are 
 \begin{equation}
 	\varepsilon_{\pm, j}^\tau=\pm \sqrt{\left(\hbar v_F \kappa_j^\tau\right)^2+\frac{\gamma_1^2}{2}+\tau \sqrt{\left(\hbar v_F \kappa_j^\tau\right)^2\gamma_1^2+\frac{\gamma_1^4}{4}}},
 	\label{Eq31}
 \end{equation}
 where  the wave vector is
 \begin{align}
 \kappa_j^\tau=\sqrt{(k_j^\tau)^2 +\left(k_y+\frac{d_j}{l_B^2}\right)^2}\label{Eq17}	
 \end{align}
 $k_1^\tau$ represent the wave vectors of the propagating wave in the first region, where there are two right-going (incident) propagating modes and two left-going (reflected) propagating modes. Similarly, $k_3^\tau$ denote the wave vectors of the propagating wave in the third region, featuring two right-going (transmission) propagating modes.
%
\section{Transport properties}\label{Trans}

To determine the transmission probability, in the first step, we apply continuity conditions at each interface of the present system. In the second step, we use the transfer matrix method to establish a link between the coefficients $\mathcal{C}_{\text{1}}$ in the incident region and $\mathcal{C}_{\text{3}}$ in the transmitted region. From \eqref{Eq266}, we deduce
%
\begin{subequations}
        \begin{align}
&\mathcal{C}_{\text{1}}=\left(
        \delta_{\tau, 1},
        r_{+}^{\tau} ,
        \delta_{\tau,-1} ,
        r_{-}^{\tau}
        \right)^{\dagger},\\
&\mathcal{C}_{\text{3}}=\left(
        t_{+}^{\tau}
       ,0,
        t_{-}^{\tau},
        0\right)^{\dagger},
        \end{align}
\end{subequations}
where $\delta_{\tau,\pm1}$ denotes the Kronecker delta symbol. Then, at interfaces $x=d_1$ and $x=d_2$, the continuity of eigenspinors \eqref{Eq8} and \eqref{Eq24} allows to write 
\begin{subequations}
\begin{align}
&
\mathcal{L}_{1} \mathcal{M}_1(d_1) \mathcal{C}_{1}=\mathcal{L}_{2} \mathcal{M}_{2}(d_1) \mathcal{C}_{2} ,
\label{Eq33}\\
&
\mathcal{L}_{2} \mathcal{M}_{2}(d_2) \mathcal{C}_{2}=\mathcal{L}_{3} \mathcal{M}_{3}(d_2) \mathcal{C}_{3}.
\label{E34}
\end{align}
\end{subequations}
After some algebra, we establish the relation 
\begin{equation}
\mathcal{C}_{\text{1}}=\mathcal{N} \mathcal{C}_{\text{3}},
\label{Eq20}
\end{equation}
where the transfer matrix $\mathcal{N}$ is defined as
\begin{equation}
\mathcal{N}=\prod_{j=\text{1}}^{\text{2}} \mathcal{M}_{j}^{-1}(d_{j}) \mathcal{L}_{j}^{-1} \mathcal{L}_{j+1} \mathcal{M}_{j+1}(d_{j}).
\label{E38}
\end{equation}
One can use Eq. (\ref{Eq20}) to derive the following transmissions coefficients 
\begin{align}
&	t_{+}^{\tau}=\frac{\mathcal{N}_{33} \delta_{\tau, 1}-\mathcal{N}_{13} \delta_{\tau,-1}}{\mathcal{N}_{11} \mathcal{N}_{33}-\mathcal{N}_{13} \mathcal{N}_{31}},
\\
&	t_{-}^{\tau}=\frac{\mathcal{N}_{11} \delta_{\tau,-1}-\mathcal{N}_{31} \delta_{\tau, 1}}{\mathcal{N}_{11} \mathcal{N}_{33}-\mathcal{N}_{13} \mathcal{N}_{31}},
\end{align}
where $\mathcal{N}_{ij}$ are matrix elements of $\mathcal{N} $ \eqref{E38}. To completely determine the corresponding transmission probabilities, we introduce  the current density 
$\textbf{j}=v_{F}{\Psi}^{\dagger}\vec{\alpha}\Psi$ associated with the present system, and  $\vec\alpha$ is a $4\times4$ matrix with two Pauli matrices $\sigma_{x}$ on its diagonal. Then we use the  incident $\textbf{j}_{\text{inc}}$ and transmitted $\textbf{j}_{\text{tra}}$    current densities
\begin{align}
    T_{\pm}^{\tau}
    =\frac{\left|\textbf{j}_{\text{tra}}\right|}{\left|\textbf{j}_{\text{inc}}\right|}
    =\frac{k_{3}^{\tau}}{k_{1}^{\tau}}\left|t_{\pm}^{\tau}\right|^{2}.
\label{E43}
\end{align}

Concerning the conductance, we apply the Landauer-Büttiker formula \cite{Blanter} at zero temperature to derive
\begin{equation}
\mathrm{G}(E)=G_0 \frac{W}{2 \pi} \int_{-\infty}^{+\infty} \mathrm{d} k_y \sum_{\tau, n= \pm} T_n^\tau\left(E, k_y\right),
\end{equation}
where $G_0=4e^2/h$, the factor 4 count the valley and spin degeneracy in graphene and $W$ is the width of the sample along the $y$-direction. The conductivity is then expressed as \cite{Snyman}
\begin{equation}
\sigma=\frac{L}{W}G(E).
\end{equation}
Next, we will focus on the numerical analysis of the results obtained so far. For convenience, we will adopt the configuration of different parameters: $\vartheta_{0}/\gamma_{1}=l/l_B$ and $k_y \equiv l k_y$, with $l=\hbar v_F/\gamma_1=1.76$nm.

\section{Results and discussions}\label{RRDD}
\begin{figure*}[tbh]
        \begin{center}
        \end{center}
        \includegraphics[width=7.2in]{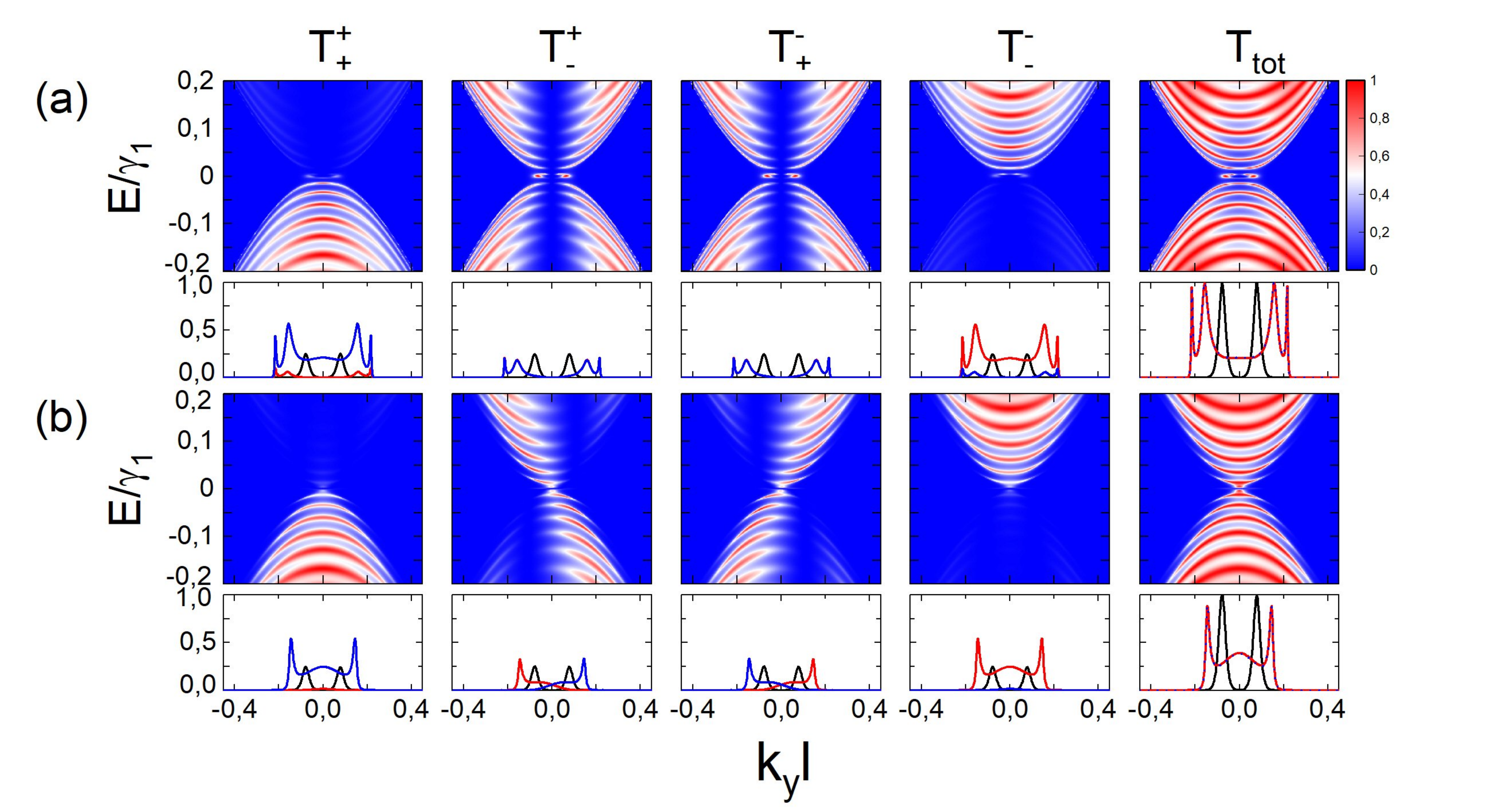}
        \caption{(Color online) Density plot of transmission probabilities as a function of energy $E$ and transverse wave vector $k_y$ for $U_\infty=-50\gamma_1$, $L=d_2-d_1=50l$, and two values of magnetic field (a): $B=0$ and (b): $B=3.61$T. The line plots correspond to the transmission at fixed energy: $E=-0.05\gamma_1$ (blue), $E=0$ (black), and $E=0.05\gamma_1$ (red).
        }
        \label{Transmission}
\end{figure*}
Fig.~\ref{Transmission} presents a comparison of the transmission probabilities under two magnetic field conditions: (a) $B=0$ and (b) $B=3.61$T. It is evident that the transmissions are broader for $B=0$ and narrower for $B=3.61$T. In Fig.~\ref{Transmission}b, the channels $T^+_+$ and $T^-_-$ demonstrate that the presence of $B$ decreases the number of resonances compared to the corresponding channels for $B=0$, see Fig.~\ref{Transmission}a. Additionally, there is an electron-hole asymmetry because we have $T^+_+(E) \neq T^+_+(-E)$ and $T^-_-(E) \neq T^-_-(-E)$ for both values $B=0$ and $B=3.61$T, as reflected in the difference between the blue and red lines. This result aligns with that derived by applying an in-plane pseudomagnetic field to AB-BLG \cite{Abdullah}. However, the correspondence  $T^+_+(-E)=T^-_-(E)$ is hold (see equivalence between blue line in $T^+_+$ and red line in $T^-_-$). 
On the other hand, in channels $T^+_-$ and $T^-_+$  for $B=0$, the transmissions exhibit electron-hole symmetry, as indicated by the matching in blue and red lines. 
Contrarily, the introduction of $B$ breaks such symmetry and  that   associated with normal incidence ($k_y=0$) in $T^+_-$ and $T^-_+$, as clearly depicted by the distinct behavior of the blue and red lines in Fig. \ref{Transmission}b. This contrasts with the result obtained in \cite{Abdullah}.
It is worthily noted that $T^+_-$ and $T^-_+$ are equivalent under the sign change of the transverse wave vector  $k_y$, i.e., $T^+_-(k_y)=T^-_+(-k_y)$. This is in agreement with the findings observed for diffusive transports when a bias is applied \cite{Van,Nadia}. 
Fig. \ref{Transmission}a shows that the channels $T^+_-$ and $T^-_+$ exhibit anti-Klein tunneling at normal and near-normal incidence for $B=0$, similar to the results for diffusive transport \cite{Van,Hassane}. 
At $B=3.61$T, transmissions take place at normal incidence, and anti-Klein tunneling is shifted towards the side.
As for total transmission $T_{tot}=\sum_{\tau,n=\pm}  T_n^\tau$, the electron-hole symmetry and symmetry with respect to $k_y=0$ are present for both values $B=0$ and $B=3.61$T. In other words, we have the symmetry $T^+_+(-E)=T^-_-(E)$ for all $B$, but  $T^+_-(E,k_y)=T_+^-(E,k_y)$ holds for $B=0$ and   $T^+_-(E,k_y)=T^-_+(-E,-k_y)$ is valid for non-null $B$, i.e., $B=3.61$T.

\begin{figure}[tbh]
\centering
\includegraphics[width=\linewidth]{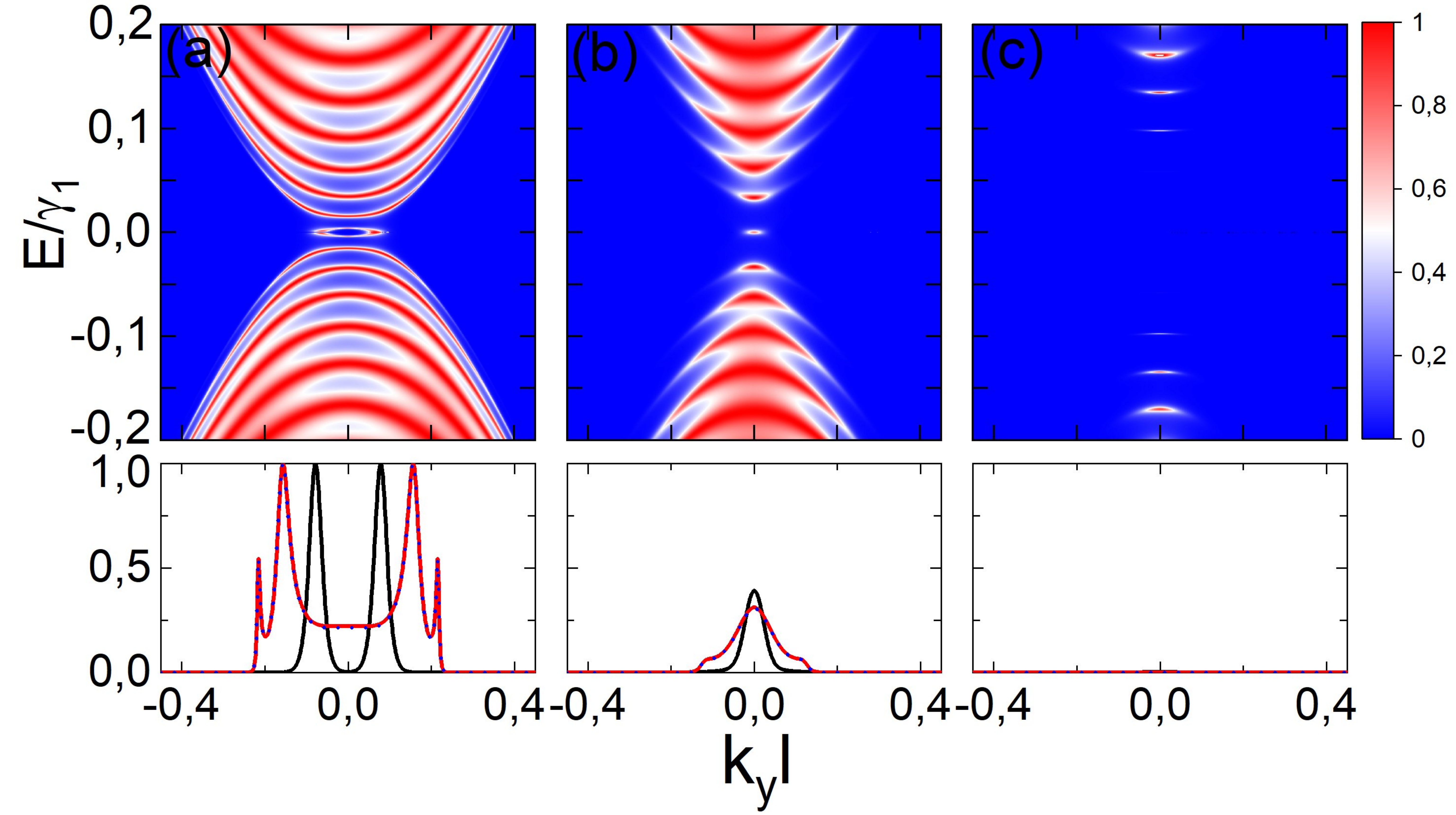}
\caption{(Color online) Density plot of the total transmission as a function of energy $E$ and  transverse wave vector $k_y$ with the same parameters as in Fig. \ref{Transmission} but for (a): $B=0.9$ T, (b): $B=7.37$ T, and (c): $B=14.44$T.}\label{TTransmission}
\end{figure}

Fig. \ref{TTransmission} shows the total transmission $T_{tot}$ for three different values of the magnetic field (a): $B=0.9$T, (b): $B=7.37$T, and (c): $B=14.44$T. It is observed that as we increase  $B$, the transmission narrows towards the center, and then anti-Klein tunneling increases. In addition, in Fig. \ref{TTransmission}b, the number of resonances decreases while their thickness increases compared to Fig. \ref{TTransmission}a. Moreover, two regions of zero transmission are found on either side of the neutrality point ($E=0$). Furthermore, in Fig. \ref{TTransmission}c, the application of a high magnetic field ($B=14.44$T) results in destructing tunneling effect.
\begin{figure}[tbh]
\centering
\includegraphics[width=\linewidth]{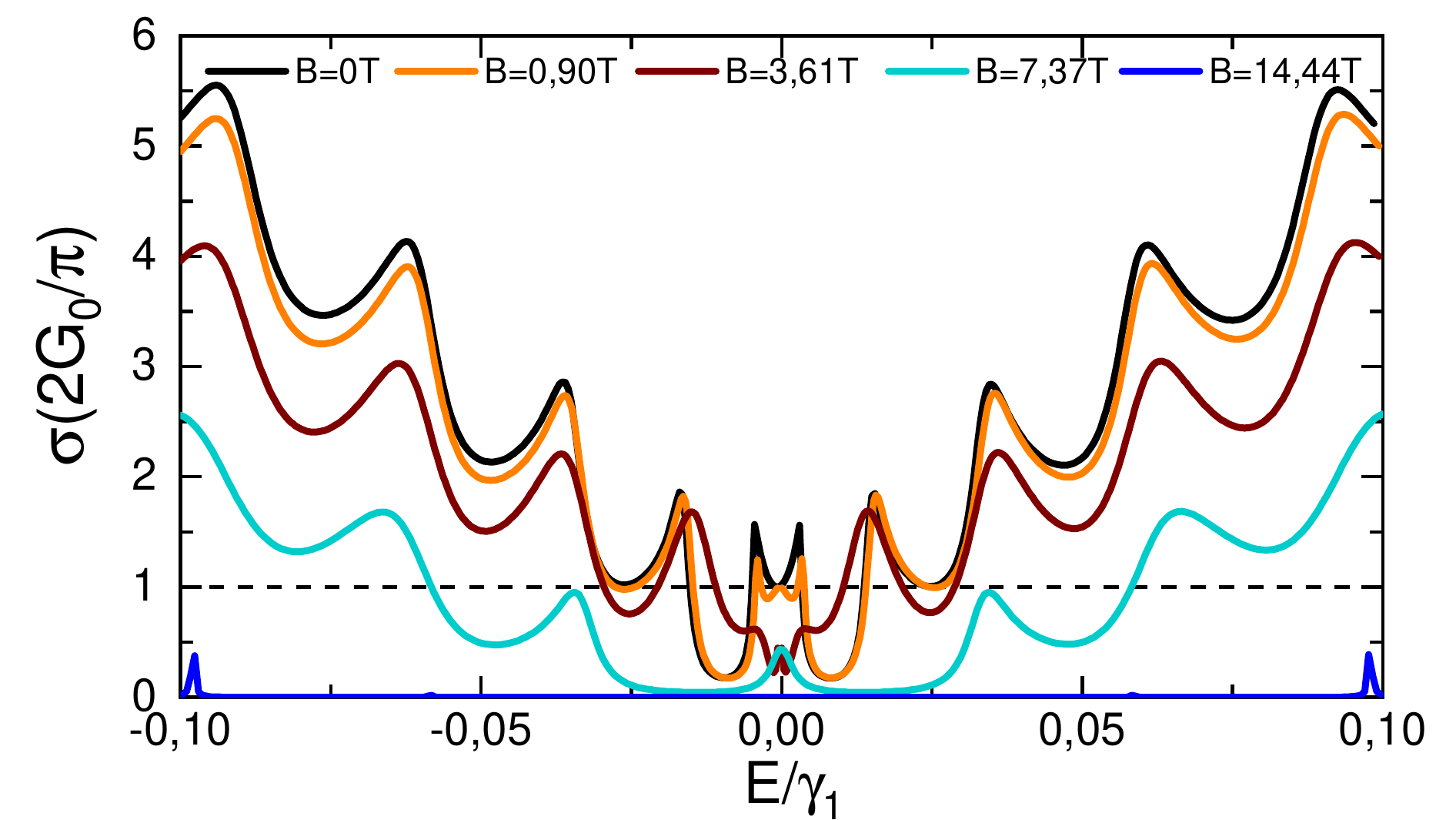}
\caption{(Color online) Conductivity as a function of energy $E$, corresponding to the transmissions $T_{tot}$ in Figs. \ref{Transmission} and \ref{TTransmission}.}\label{Conductivity}
\end{figure}

In Fig. \ref{Conductivity}, we plot the conductivity corresponding to each transmission $T_{tot}$ in Figs. \ref{Transmission} and \ref{TTransmission}. For $B=0$ (black line), the conductivity is equivalent to the result found in  \cite{Snyman,Abdullah}, with $\sigma_0=2 G_0 / \pi$ and the subscript 0 in $\sigma$ refers to  $B=0$ at the neutrality point ($E=0$).
For $B=0.9$T (depicted by the orange line), the conductivity experiences a slight decrease but maintains the same number of peaks as observed for $B=0$, reaching the value $\sigma=0.98 \sigma_0$ at the neutrality point. 
For $B=3.61$T (depicted by the brown line), the conductivity experiences a decrease, with a simultaneous reduction in the number and amplitude of the oscillations. We also observe that its value at the neutrality point drops to $\sigma=0.45 \sigma_0$.
Furthermore, with an increase in the magnetic field to $B=7.37$T (represented by the cyan line), two regions of zero conductivity become apparent, accompanied by a further reduction in the number and amplitude of oscillations compared to the results for $B=3.61$T (brown line).
 Nevertheless, the conductivity remains non-zero at the neutrality point and takes the value $\sigma=0.43\sigma_0$.
 At $B=14.44$T (indicated by the blue line), the conductivity drops to zero even at the neutrality point but exhibits two peaks at the extremities $ E =\pm0.1 \gamma_{1}$. It is evident that, as the magnetic field increases, the conductivity decreases, even at the neutrality point. This is in contrast with the result for the case of an in-plane pseudomagnetic field \cite{Abdullah}. Additionally, it is worth noting that, in all cases, the conductivity is symmetric with respect to the neutrality point due to the symmetry in the transmission probability $T_{\text{tot}}$.

\begin{figure}[tbh]
\centering
\includegraphics[width=\linewidth]{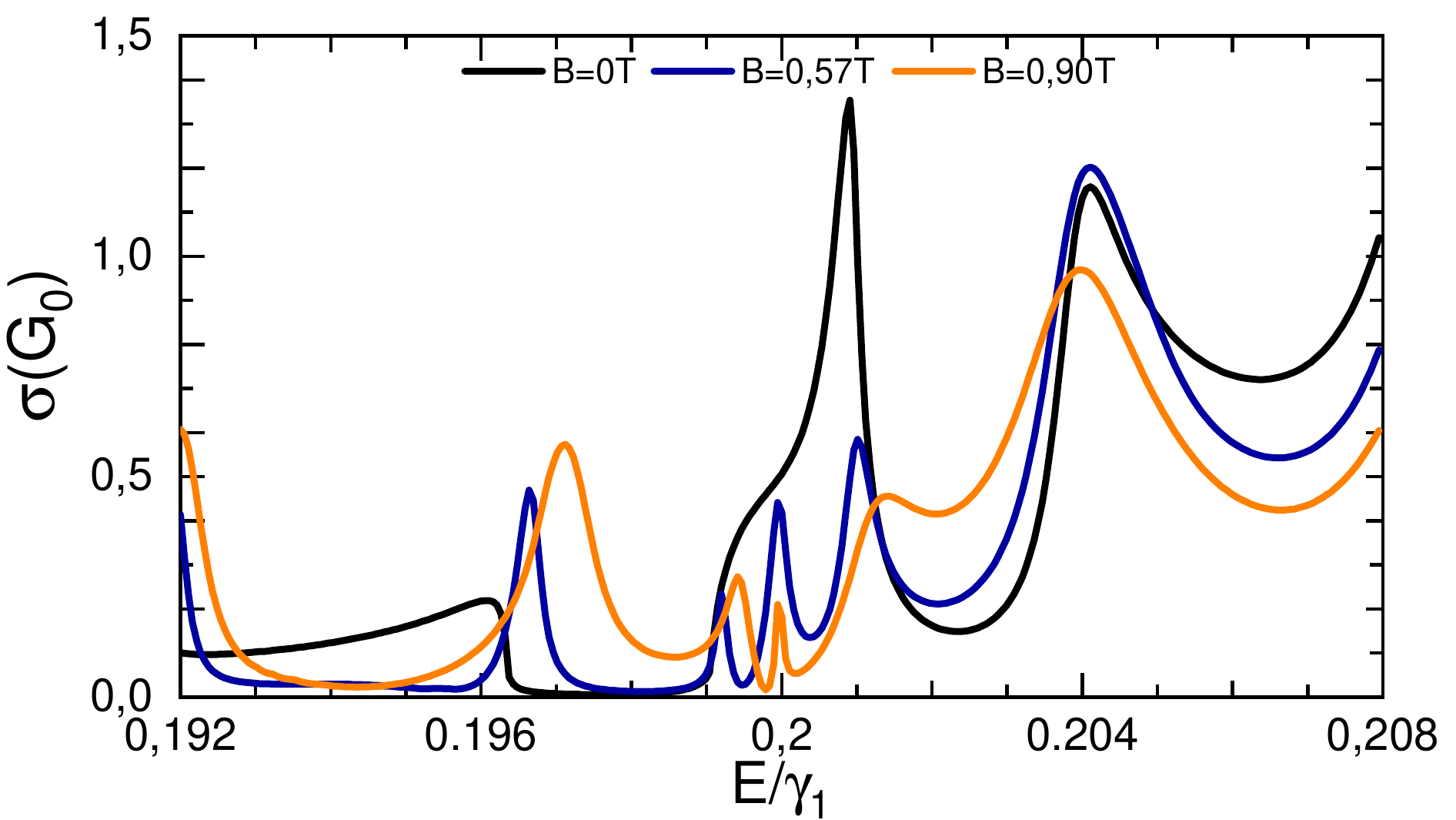}
\caption{(Color online) Conductivity as a function of energy $E$ around  Dirac point for $U_{\infty}=0.2\gamma_1$ and $L=100l$.}\label{Conductivity02U}
\end{figure}

Fig. \ref{Conductivity02U} shows the conductivity as a function of energy $E$ for three values of the magnetic field: $B=0$ (black line), $0.57$T (blue line), and $0.9$T (orange line). We reduce the potential $U_\infty$  to $0.2$$\gamma_{1}$ and choose $L=100l$. We observe that the conductivity shows an asymmetry behavior with respect to the Dirac point $E=U_{\infty}$. This asymmetry persists irrespective of the presence or absence of $B$. However, the conductivity behaves differently depending on the variation of $B$. Indeed, the conductivity displays oscillations with an increasing amplitude as $B$ increases for the case $E < U_{\infty}$, and conversely, it decreases with an increase of $B$ for $E > U_{\infty}$. In addition, at $E=U_{\infty}$, the conductivity decreases with an increase of $B$, similarly to the result presented in Fig. \ref{Conductivity} at $E=0$.

\section{Conclusion}\label{CC}

We have investigated the impact of a perpendicular magnetic field on the ballistic transport properties of AB-BLG graphene, as well as weakly doping the system. Our results have revealed several important findings. Firstly, we have observed that the presence of a magnetic field leads to a narrowing of the transmissions, a decrease in the number of resonances, and an increase in anti-Klein tunneling. Notably, we have identified electron-hole asymmetry in the channels $T^+_+$ and $T^-_-$, regardless of the presence or absence of a magnetic field. However, in the channels $T^+_-$ and $T^-_+$, we have discovered that electron-hole symmetry holds in the absence of a magnetic field, but is broken in its presence. Additionally, the symmetry with respect to $k_y=0$ is also broken. Nevertheless, due to the equivalence between these channels under the sign change of energy $E$ or transverse wave vector $k_y$, the transmission remains symmetric in $T_{tot}$. Furthermore, we have observed the destruction of  tunneling effect as a result of considering a high magnetic field.

In the context of ballistic transport, we have found that conductivity decreases with an increasing magnetic field. However, in the case of a weakly doped system with a potential of $U_{\infty}=0.2\gamma_1$, we have observed an asymmetry in the conductivity at the Dirac point $E=U_{\infty}$. Additionally, we have noted that the amplitude of oscillations increases with the magnetic field for $E<U_{\infty}$ but decreases in the opposite case, $E>U_\infty$.
In conclusion, our findings highlight the ability to regulate electron transmission and conductivity in AB-BLG by applying a perpendicular magnetic field in the context of ballistic transport. These results provide valuable insights into the behavior of AB-BLG graphene under the influence of a magnetic field and weak doping, which can contribute to the development of graphene-based electronic devices with tailored transport properties.

\end{document}